# A Comprehensive Study of the Current State-of-the-Art in Nepali Automatic Speech Recognition Systems

Rupak Raj Ghimire*, Bal Krishna Bal† and Prakash Poudyal‡
Information and Language Processing Research Lab
Kathmandu University, Nepal
Email: *rughimire@gmail.com, †bal@ku.edu.np, ‡prakash@ku.edu.np

*Abstract*—In this paper, we examine the research conducted in the field of Nepali Automatic Speech Recognition (ASR). The primary objective of this survey is to conduct a comprehensive review of the works on Nepali Automatic Speech Recognition Systems completed to date, explore the different datasets used, examine the technology utilized, and take account of the obstacles encountered in implementing the Nepali ASR system.

In tandem with the global trends of ever-increasing research on speech recognition based research, the number of Nepalese ASR-related projects are also growing. Nevertheless, the investigation of language and acoustic models of the Nepali language has not received adequate attention compared to languages that possess ample resources. In this context, we provide a framework as well as directions for future investigations.

## I. INTRODUCTION

Nepali is the official language of Nepal and belongs to the Indo-Aryan language family. It is spoken by a significant number of people in Nepal, Bhutan, and India. The Nepali language is traditionally written in the Devanagari script, which originated from the Brahmi script during the 11[th] century. It has 11 vowels (अ,आ,इ,ई,उ,ऊ,ऋ,ए,ऐ,ओ,औ) , 33 consonants (क,ख,ग,घ,ङ, च,छ,ज,झ,ञ, ट,ठ,ड,ढ,ण, त,थ,द,ध,न, प,फ,ब,भ,म, य,र,ल,व,श,स,ष,ह ), 7 vowel modifiers (ा, ी, ि, ु, ू, ो, ै, ो, ौ), numerals (०, १, २, ३, ४, ५, ६, ७, ८, ९ ), and other symbols such as ं, ः, ँ [1].There are some complex consonants such as क्ष, त्र, ज्ञ which are the combination of other consonant and markers (Example: क्ष = क + ् + ष). The Nepali language has limited computational resources causing it to significantly lag behind in the development of language technologies compared to resourceful languages like English, French, and German.

Automatic Speech Recognition (ASR) is a cutting-edge technology that enables computers to comprehend and interpret human speech. It is widely recognized as a pivotal tool in the field of Human-Computer Interaction. ASR systems serve a wide array of application domains, for example, virtual assistant technology, call center automation, speaker identification, medical applications, etc. ASR systems are exhibiting a notable trend towards enhanced accuracy and reliability. However, several obstacles are still prevalent that require the researchers' attention like the presence of noise in input data, the inadequacy of current systems to handle various accents and dialects, the lack of effective means to address side talks and multiple speakers speaking simultaneously, etc. Meeting the basic level of accuracy for low-resourced languages like Nepali is even more challenging.

The classical ASR pipeline involves multiple phases including Feature Extraction, Acoustic Modeling, Language Modeling, and Decoding. In contrast, modern ASR systems referred to as End-to-End ASR models, integrate all stages into a single Deep Learning model.

Nepali is regarded as a low-resource language because of the availability of very less computational resources which are necessary for the development of language technologies. Despite of the availability of less amount of resources, there are a couple of works [2][3][4] done on the Natural Language Processing (NLP) domain including the works on text and speech processing. The initiation of language processing works started with the PAN localization and the Bhasha Sanchar projects led by Madan Puraskar Pustakalaya (MPP). Through these projects, Nepali National Corpus comprising of Monolingual Nepali Written Corpus, English to Nepali Parallel Corpus, Spell-checker, Text-To-Speech (TTS), Optical Character Recognition (OCR), and Part-Of-Speech Tagging tools and systems were developed[1]. Comparatively, works on Speech synthesis and recognition were not that extensively worked upon within the scope of these projects.

The evaluation of ASR systems commonly relies on the utilization of two widely accepted metrics, namely Word Error Rate (WER) and Character Error Rate (CER) [5]. Both of these metrics serve as effective means to assess the accuracy of ASR systems.

This paper is structured in the following manner: Sec-tion II provides a concise overview of the methodology employed to identify and obtain the relevant scholarly literature. Section III provides an explanation of the Nepali speech corpus that is currently accessible. Similarly, section IV discusses the acoustic and language models that have been covered by the literature selected for the purposes of this review. Following a concise assessment of the Nepali ASR systems in section V, we then discuss the challenges that were encountered, potential avenues for further investigation, and the practical domains where the Nepali ASR can be applied in section VI. We conclude the paper in Section VII.

---

[1]Projects by Language Technology Kendra: https://ltk.org.np/projects.php

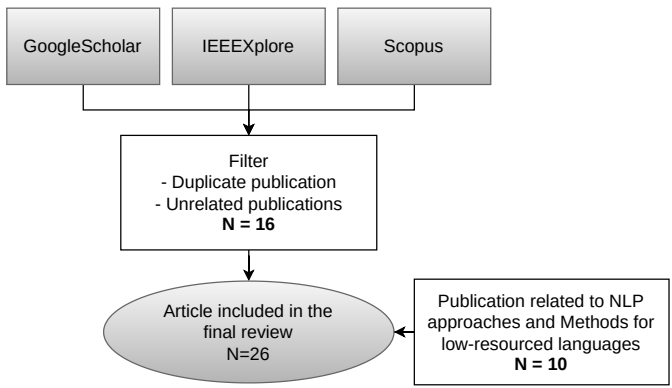

Figure 1. Procedure to review the publications reviewed in this survey

## II. Selection of the Literature

In order to determine the article for review, a preliminary step involved the selection of keywords such as "Nepali ASR", "Nepali NLP", and "Nepali Speech Recognition System" among others. A survey was conducted on widely-used databases, namely Google Scholar, IEEE Xplore, and Scopus. The comprehensive procedure is illustrated in Figure 1. Upon conducting a thorough search, we proceeded to evaluate the search results by considering the name of the publication and relevant keywords. In the process of selecting a limited number of candidates, we thoroughly examined the abstracts and applied a filtering mechanism to identify publications that were pertinent to our research. As a result, a total of 16 relevant literature sources were identified within the scope of this study. In addition to the aforementioned literature, we have also incorporated supplementary publications that encompass research conducted in the context of low resource settings.

## III. Nepali Speech Corpus

The publication of the Nepali National Corpus (NNC)[6] as a part of the Bhasha Sanchar Project[1] marked a significant milestone in the field of computational language resource development for the Nepali language. The Nepali National Corpus (NNC) consists of three distinct corpora, 1) Nepali Monolingual Written Corpus 2) English-Nepali Parallel Corpus, and 3) Nepali Spoken Corpus. The Nepali Spoken Corpus distinguishes itself as the inaugural corpus of its kind to be published within the academic realm. This corpus consists of a total of 260,000 words and 6,000 recorded sentences. The resources can be accessed from the ELRA website[2]. Several Nepali Text-to-Speech(TTS) systems [7][8] have been developed based on this corpus. Allwood et al. [9] conducted an activity-based corpus development to caputure the authentic context accurately. According to the authors, a total of 61 hours of recordings were generated, documenting various activities. The data collection was carried out at a community level, with the assumption that the real-world scenario could be effectively captured within this particular setting.

Over time, independent research publications [10][3][11] have claimed that they prepared the data source and conducted experiments based on those datasets. However, it should be noted that none of these sources have been published or made available for open access in the public domain. According to Deka et al. (2018)[12], preparations are underway to develop a speech dataset for the Nepali and other languages spoken in the northeastern region of India. Unfortunately, the authors report that the ongoing work is yet to be completed, and at the same time we are currently unable to locate any of their subsequent publications or resources in the public domain.

Kjartansson et al. [13] can be considered a significant achievement in the development of Nepali speech corpora especially for the ASR system development.A total of 157, 905 recordings yielding 165 hours of speech data were collected from 527 distinct speakers. Similarly, Sodimana et al. [14] have prepared Text-to-Speech (TTS) dataset. These datasets are accessible via the Open Speech and Language Resource (OSLR) website[3] [4]. With the release of these datasets, numerous research studies[4][2][15] have utilized them as a primary referral of the data source.

## IV. Methodologies of Nepali ASR

ASR systems are more complex NLP systems compared to text-based systems. In order to develop the ASR system, it is essential to employ two distinct models, namely the acoustic model and the language model. This section provides an explanation about how those models are utilized in the context of the Nepali ASR system.

### A. Acoustic Models

The Hidden Markup Model (HMM) and Gaussian Mixture Models (GMM) are recognized as the predominant statistical methods for modeling the sequential nature of speech processing tasks [16][17][18]. The implementation of HMM-based systems is comparatively simple and can be accomplished with a smaller training dataset. Early stage ASR and TTS systems are developed using HMM. The advent of Artificial Neural Network (ANN) and supervised Machine Learning (ML) techniques, particularly Deep Neural Networks (DNN) and associated algorithms, have propelled progress in speech processing research. The Mel-frequency Cepstral Coefficients (MFCC) or Mel-Spectrogram are commonly regarded as the acoustic input feature vector for the HMM and DNN-based ASR modules. Likewise, the Convolution Neural Network (CNN) captures the local patterns and characteristics present in speech signals[19]. The temporal dependencies present in the speech signal, both in the short-term and long-term, can be effectively captured through the utilization of specialized forms of Recurrent Neural Networks (RNN) known as long-short-term memory (LSTM)[20] and Gated Recurrent Unit (GRU)[21]. The Transformer model is widely regarded as a cutting-edge model and is extensively utilized in the field of end-to-end (E2E) models owing to its capacity to efficiently capture long-term dependencies in speech signals. The Connectionist Temporal Classification (CTC) is the way to align speech with the sequence of the written text[19]. A CTC-based loss function allows for training DNN-based E2E ASR models.

---

[2]ELRA: http://www.elra.info/en/catalogues/free-resources/nepali-corpora/
[3]OSLR54: Open Speech and Language Resource: https://www.openslr.org/54/
[4]OSLR43: Open Speech and Language Resource: https://www.openslr.org/43/

The initial exploration of the Nepali ASR system was conducted in 2017, utilizing a HMM based isolated word model by Ssarma et al. [11]. The Voice Activity Detection (VAD) mechanism was used to separate the word from the speech. The Hidden Markov Model (HMM) was employed to train the model, resulting in an accuracy of 74.99% for single-word inputs and 55.55% for three-word phrase inputs. Regmi et al. [10] examined RNN-CTC based model on self-prepared Nepali speech dataset. The CTC loss function and beam search were used for the purpose of training and decoding technique respectively. The authors also experimented effect of the n-gram language model on RNN-CTC based network. Using the language model the CER improved from 42% to 34% on the same speaker, whereas CER decreased to 52% when the new and unknown speakers speech is given as input.

Baral and Shrestha [22] used Open SLR dataset (OSLR54)[13] for experimenting the different ASR models, using DNN and HNN. The study focuses on enhancing the input feature space of HMM and DMM models. Speaker adaptation is a crucial aspect in speech recognition systems, and it involves the application of different techniques. These techniques include Maximum Likelihood linear transform (MLLT), Speaker Adaptive Training (SAT), and Feature-based Maximum likelihood linear regression (fMLLR). Linear Discriminant Analysis (LDA) is a technique employed to decrease the dimensionality and computational expenses. The dynamic speech information is captured by delta-cepstral features, which are obtained through the utilization of delta and delta-delta optimization techniques. The experiment demonstrated that the STA-LDA-MLLT training method yielded the lowest word error rate (WER) of 29.45% when applied to 150 hours of speech data. In contrast, the Delta-Delta and LDA-MLLT models achieved WERs of 39.15% and 33.59% respectively. A dataset consisting of 200 hours of data has been utilized for the DNN models. The study introduces three deep neural network (DNN) based models, namely: 1) $DNN-sMBR$, 2) $DNN-Pnorm$, and 3) $DNN-TDNN-LSTM$. The $DNN-TDNN-LSTM$ based model demonstrates greater accuracy compared to all other models, getting a WER of 11.55% (refer to Table I for further details on the experiment).

Bhatta et al. [4] conducted an experiment to develop a model based on CNN-GRU-CTC for Nepali speech recognition. The authors utilized a high-quality text-to-speech (TTS) dataset obtained from OSLR43 [14]. The model with Convolutional Neural Network (CNN) proposed in this study demonstrated superior performance compared to the RNN-CTC model as suggested by Baral and Shrestha [22]. Banjara et al. [23] conducted a comparative experiment on different RNN based ASR models. The experiment involved direct input feeding to various RNN architectures, including Gated Recurrent Unit (GRU), Long Short-Term Memory (LSTM), Bidirectional LSTM (BiLSTM), and one-dimensional ($1D-CNN$), as well as combinations of CNN with the aforementioned RNN architectures. It has been observed that the combination of CNN and RNN exhibits strong performance and yields better output along with CTC beam search decoding. The experiment reported 27.72% CER.

The authors Bhatta et al. [4] and Banjara et al. [23] both proposed a similar architecture. However, there is a significant difference in the CER i.e. 1.84% and 27.72% respectively. In contrast to the similar architectural framework Bhatta et al. [4] used high quality TTS dataset (OSLR43[14]) and Banjara et al. [23] used ASR dataset (OSLR54[13]). This is one of the reasons behind the significant difference in the CER. The variation in outcomes observed in the same model can be attributed to the utilization of distinct datasets and varying sampling rates during the generation of feature vectors (MFCCs).

Regmi and Bal [2] introduced the E2E Nepali ASR model, which combines the CTC and Attention mechanisms. In their study, the model is trained using a combined dataset from OSLR54 [13], OSLR43[14], and a 167 Minute additional dataset prepared by themselves. The CER measured is 10.3%. This is the best CER among all the proposed Nepali ASR models that were trained using the OSLR54 dataset.

Dhakal et al. [24] introduced a model that uses a combination of Residual Network (ResNet) within BiLSTM and 1D-CNN framework trained using CTC loss function. The ResNet architecture is employed to enhance the efficiency of network training and mitigate premature saturation of the CTC-loss function. As part of the comparative analysis, the researcher evaluated the performance of several models, namely $1D-CNN+BiLSTM$, $1D-CNN+ResNet$, $1D-CNN+ResNet+BiGRU$, $1D-CNN+$

$ResNet + BiLSTM$, and $CNN + ResNet + LSTM$. Among these models, it was found that the combination of 1D-CNN, ResNet, and BiLSTM yielded better results, achieving a CER of 17.06%. This finding demonstrates that the addition of ResNet has the potential to greatly enhance the performance of the $1D-CNN+GRU$ model as recommended by Banjara et al. [23] in their work. Joshi et al. [15] conducted experiment on the CNN and RNN based architecture. According to the experiment, it has been determined that the architecture that uses the E2E model with the combination of $CNN+BN+GRU+BN+Dense\ Layer+Softmax+CTC$ exhibits the best results. This architecture achieved a WER of 10.67% on the TTS dataset (OSLR43[14]) which is prepared in controlled indoor environment free of any noise interference.

Joshi et al. [3] conducted an experiment on an ASR system using a dataset they created by themselves, consisting of 23 hours of audio data. The dataset is augumented to a duration of 80 hours for the purpose of training. A number of experiments were conducted on different models namely $CNN+1-GRU$, $CNN+2-GRU$, and $CNN+3-GRU$. The results obtained by the researcher indicate that the $2-GRU$ model exhibits faster computational performance when compared to the other models. It has been observed that increasing the number of GRU units from 200 to 400 lead to an improvement in the model's performance in terms training loss. However, this enhancement came at the cost of increased training time. The author also applied a uni-gram language model and observed that the implementation of the language model resulted in an improvement in the WER from 49.30% to 37.50%. This model exhibits similarities with the work of Banjara et al. [23]. However, it uses a different evaluation metric, making direct comparisons between the two models unfeasible.

The task of gender classification based on voice has been introduced for Nepali speech data by Danuwar et al. [25], who utilized MFCC in conjunction with GMM. The aforementioned concept has been put forth as a potential voice bio-metric retrieval system. On the test dataset accuracy rate of 93.8% has been reported.

### B. Language Models

To enhance the precision of automatic speech recognition (ASR) systems, language-specific models can be employed. These models have the capability to rectify predictions, thereby leading to an improvement in accuracy. N-Gram models are extensively employed as language models that exhibit independence from any specific language. N-gram models possess the capability to take into account the preceding n words with the help of greedy or beam search based decoding approach. Transformer-based language models have gained popularity as a means of enhancing the overall performance of deep neural network (DNN) models by rectifying their output.

The utilization and implementation of n-gram, as a basic statistical model, is comparatively simpler within the context of training and integrating it into an end-to-end (E2E) model. Regmi et al. [10] applied an n-gram based language model on proposed architecture. Their approach resulted in a reduction of the CER from 42% to 34%. Joshi et al. [3] also used an uni-gram language model and observed that the use of the language model resulted in an reduction of WER from 49.39% to 37.50%. Similarly, the study conducted by

Regmi and Bal [2] employed a language model based on recurrent neural networks- a one-layer LSTM RNN-LM. This addition resulted in the increment of the performance compared to all other Nepali ASR models on a multi-speaker dataset such as OSLR54.

## V. EVALUATION OF NEPALI ASR

The evaluation of the Nepali ASR model in existing research predominantly employs WER and/or CER as metrics. The research conducted is elaborated upon in Section IV-A and summarized briefly in Table I.

According to the analysis, it can be concluded that the study conducted by Bhatta et al. [4] demonstrates the lowest CER among all the examined cases, with a value of 1.84%. The training and evaluation was done based on the single-speaker high-quality speech dataset designed for TTS[14]. However, the model proposed by Regmi and Bal [2] achieved a CER of 10.30% on the multi-speaker dataset designed for ASR[13], and the model proposed by Joshi et al. [15] achieved WER of 10.67% on the high quality TTS dataset[14]. These results are currently regarded as the state-of-the-art in the domain of Nepali speech recognition[5].

## VI. DISCUSSION

The data presented in Table I indicates a noticeable rise in Nepali ASR research efforts in recent years, in line with the global trends. The state-of-the-art ASR model has been applied within the development for the Nepali ASR model. However, the output of the Nepali ASR model does not yet compare with the performance of models designed for languages with abundant resources. The observations made during the review can be summarized in the following subsections:

### A. Challenges in building ASR for Nepali

*1) Dataset for Nepali ASR:* The advancement of industry-standard ASR requires the acquisition of substantial quantities of high quality speech data. There exists a limited number of publicly accessible speech datasets, which are currently being utilized by researchers. The publicly available dataset suffers from several shortcomings such as the presence of noise in the recordings, long pauses, and in general low quality. An additional

---

[5]This information is based on the latest literature review done untill July 26, 2023

Table I
Proposed Nepali ASR Models and their evaluation

| Paper | Author, Year | Data Set | Algorithm / Model / Techniques | Metrics | Result |
|---|---|---|---|---|---|
| Nepali Speech Recognition using CNN, GRU and CTC | Bhatta et al., 2020 | OSLR43 | CNN-GRU-CTC | CER | 1.84% |
| An End-to-End Speech Recognition for the Nepali Language | Regmi and Bal, 2021 | OSLR54 + OSLR43 + *additional 167 Minutes of speech | Joint CTC.Attention | CER | 10.30% |
| Nepali Speech Recognition using CNN and Sequence Models | Banjara et al., 2020 | OSLR54 | CNN-GRU (CTC Beam search) | CER | 27.72% |
| Nepali Speech Recognition using RNN-CTC Model | Regmi et al., 2019 | * scrapped text with 1320 word and made speeh dataset | RNN-CTC (with LM and Same Speaker) | CER | 34.00% |
| | | | RNN-CTC (without LM and Same Speaker) | CER | 42.00% |
| | | | RNN-CTC (with LM and Different Speaker) | CER | 52.00% |
| Automatic Speech Recognition for the Nepali Language using CNN, Bidirectional LSTM and ResNet | Dhakal et al.. 2022 | OSLR54 | 1D-CNN+ResNet+BiLSTM | CER | 17.07% |
| | | | 1D-CNN+BiLSTM | CER | 19.71% |
| | | | 1D-CNN+ResNet | CER | 24.60% |
| | | | 1D-CNN+ResNet+BiGRU | CER | 29.60% |
| | | | CNN+ResNet+LSTM | CER | 30.27% |
| A Novel Deep Learning Based Nepali Speech Recognition | Joshi et al., 2022 | OSLR43 | CNN+BN+GRU+BN+Dense Layer+Softmax+CTC | WER | 10.67% |
| Large Vocabulary Continous Speech Recognition for Nepali Language | Baral and Shrestha, 2020 | 150 Hr OSLR54 + * additional 50 Hours of speech | | | |
| | | 200 Hours | DNN-TDNN-LSTM | WER | 11.55% |
| | | 200 Hours | DNN-sMBR | WER | 12.30% |
| | | 200 Hours | DNN-Pnorm | WER | 18.16% |
| | | 150 Hours | HMM (STA-LDA-MLLT) | WER | 29.45% |
| | | 150 Hours | HMM (LDA-MLLT) | WER | 33.56% |
| | | 150 Hours | HMM (Delta-Delta) | WER | 39.15% |
| End to End Based Nepali Speech Recognition System | Joshi et al., 2023 | * 23 Hours of Recording and 80 Hours of augumentation | CNN + 2-GRU(400 Unit) with language model | WER | 37.50% |
| | | | CNN + 2-GRU(400 Unit) without language model | WER | 49.39% |
| HMM Based Isolated Word Nepali Speech Recognition | Ssarma et al., 2017 | * | HMM | Percision | 75.00% |

*Author prepared Dataset (Not available in public domain)*

concern regarding the dataset is that it does not consider accents, speaking styles, gender representation, and age groups. In order to achieve a comprehensive general purpose ASR system, it is essential to include all aspects of spoken language within the dataset. The present dataset consists of mostly a read speech of written text, whereas for ASR the activity based speech recordings are the preferred form of data.

*2) Lack of Language Model:* The domain of supporting language models (LM) is another area that has received limited attention in in case of Nepali ASR. The inclusion of the LM is necessitated by the tonal characteristics of the Nepali language. The researchers are testing basic language models, including uni-gram, bi-gram, tri-gram, as well as primitive RNN-based models. However, it is essential to develop and also investigate other language-dependent models.

*3) Continuation of work on identified problems:* Based on the findings of the literature survey, it is found that the continuation of the research are interrupted. Conducting thorough research on the specific subject matter will facilitate the investigation of domain-related concerns and their subsequent resolution in a methodical manner.

Despite these challenges, the advancement of Automatic Speech Recognition (ASR) in the Nepali language is progressively improving due to the increased research interest in low-resourced languages. Concurrently, activities are being undertaken to collect additional speech data, foster research collaborations, and create language-specific tools and models. The further development of Nepali ASR is expected to be positively influenced by advancements in ASR technology and speech processing techniques.

### B. Possible area of work for speech technologies

Nepali ASR domain still has lots of unexplored research topics. The subsequent research directions that can be pursued are as follows:

- In order to develop a Nepali automatic speech recognition (ASR) system, researchers must allocate significant effort towards dataset collection. The collection of a dataset focused on activities (talk, debate, phone conversation etc.) should encompass the recording of all phone usage during spoken speech. This assertion is also put forth by Baral and Shrestha [22] in their study on large vocabulary continuous speech recognition.
- A significant number of Nepali speakers commonly utilize English, Hindi, and various local languages along with Nepali speech. The examination of this matter is crucial in the generation of the dataset. Another potential area of research that could be valuable is the preparation of multilingual (bi-lingual or tri-lingual) datasets [26] and the development of code-switching end-to-end models [27, 28, 29, 30, 31].
- The existing dataset has the potential to be expanded through augmentation techniques in order to enhance its size[32, 33].
- An exploratory analysis of the dataset can be conducted in order to examine the dataset's quality. The exploration of dataset cleaning and enhancing approaches for Nepali ASR can be done[34].
- Conduct an experimental investigation on the contemporary models that employ attention-based architectures, such as the Transformer and Conformer models etc.
- The Nepali Automatic Speech Recognition (ASR) system currently faces a limitation in the availability of pre-trained models for transfer learning. The model developed by scholars should be published to facilitate interested researchers in applying transfer learning and achieving improved outcomes[35, 36].

### C. Applications of Nepali ASR

The existence of industrial-level ASR applications has not yet been realized. If it is possible to develop a high-quality ASR model that can be readily implemented on Voice Assistant platforms, it would greatly facilitate the utilization of digital services by individuals lacking digital literacy skills. This technology can serve as an accessibility tool for individuals who experience speech impairments or encounter challenges in reading and writing. ASR has the potential to yield benefits in the realm of language learning and education. The implementation of domain-specific automatic speech recognition (ASR) technology has the potential to streamline and automate customer care call centers. The application of language documentation and preservation efforts can be extended to the endangered dialects of Nepali using these tools. The online resource accessibility to the people having difficulties on text based input production can be improved by developing the language tools which are voice enabled. Using these tools the digital divide can be reduced.

## VII. Conclusion

This review attempted to comprehensively examine the body of literature pertaining to the field of Nepali Automatic Speech Recognition (ASR). The model and technology employed, as well as the experimental findings have been thoroughly investigated, analyzed, and reported. Furthermore, the study also map the interconnection among the accomplished works. Finally, we conducted an in-depth examination of the prevailing patterns observed among researchers studying rich resource languages and deliberated on potential avenues for future research in the context of low-resourced languages like Nepali.

## Acknowledgement

Authors would like to extend sincere thanks to the reviewers for their constructive comments and suggestions.

## References

[1] B. K. Bal, "Structure of Nepali Grammer," in *Madan Puraskar Pustakalaya, Nepal*, 2004.

[2] S. Regmi and B. K. Bal, "An End-to-End Speech Recognition for the Nepali Language," in *18th International Conference on Natural Language Processing*, 2021, pp. 180–185.

[3] B. Joshi, B. Bhatta, and R. K. Maharjan, "End to End based Nepali Speech Recognition System," *International Journal of Signal Processing, Image Processing and Pattern Recognition*, vol. 17, no. 1, pp. 102–109, 2023.

[4] B. Bhatta, B. Joshi, and R. K. Maharjhan, "Nepali speech recognition using CNN, GRU and CTC," in *Taiwan Conference on Computational Linguistics and Speech Processing*. Conference on Computational Linguistics and Speech Processing, 2020.

[5] A. C. Morris, V. Maier, and P. D. Green, "From WER and RIL to MER and WIL: improved evaluation measures for connected speech recognition." in *INTERSPEECH*. ISCA, 2004.

[6] Y. P. Yadava, A. Hardie, R. R. Lohani, B. N. Regmi, S. Gurung, A. Gurung, T. McEnery, J. Allwood, and P. Hall, "Construction and annotation of a corpus of contemporary Nepali," *Corpora*, vol. 3, no. 2, pp. 213–225, 2008.

[7] B. K. Bal, "Towards Building Advanced Natural Language Applications - An Overview of the Existing Primary Resources and Applications in Nepali," in *Proceedings of the 7th Workshop on Asian Language Resources (ALR7)*. Suntec, Singapore: Association for Computational Linguistics, Aug. 2009, pp. 165–170.

[8] R. R. Ghimire and B. K. Bal, "Enhancing the Quality of Nepali Text-to-Speech Systems," in *Creativity in Intelligent Technologies and Data Science*, A. Kravets, M. Shcherbakov, M. Kultsova, and P. Groumpos, Eds. Springer International Publishing, 2017, vol. 754, pp. 187–197, series Title: Communications in Computer and Information Science.

[9] J. Allwood, B. N. Regmi, S. Dhakhwa, and R. K. Uranw, "An activity based spoken language corpus of Nepali," in *2012 International Conference on Speech Database and Assessments*. IEEE, 2012, pp. 24–29.

[10] P. Regmi, A. Dahal, and B. Joshi, "Nepali Speech Recognition using RNN-CTC Model," *International Journal of Computer Applications*, vol. 178, no. 31, pp. 1–6, 2019.

[11] M. K. Ssarma, A. Gajurel, A. Pokhrel, and B. Joshi, "HMM based isolated word Nepali speech recognition," in *International Conference on Machine Learning and Cybernetics (ICMLC)*. IEEE, 2017, pp. 71–76.

[12] B. Deka, J. Chakraborty, A. Dey, S. Nath, P. Sarmah, S. Nirmala, and S. Vijaya, "Speech Corpora of Under Resourced Languages of North-East India," in *Oriental COCOSDA - International Conference on Speech Database and Assessments*. IEEE, 2018, pp. 72–77.

[13] O. Kjartansson, S. Sarin, K. Pipatsrisawat, M. Jansche, and L. Ha, "Crowd-Sourced Speech Corpora for Javanese, Sundanese, Sinhala, Nepali, and Bangladeshi Bengali," in *6th Workshop on Spoken Language Technologies for Under-Resourced Languages (SLTU)*. ISCA, 2018, pp. 52–55.

[14] K. Sodimana, P. De Silva, S. Sarin, O. Kjartansson, M. Jansche, K. Pipatsrisawat, and L. Ha, "A Step-by-Step Process for Building TTS Voices Using Open Source Data and Frameworks for Bangla, Javanese, Khmer, Nepali, Sinhala, and Sundanese," in *6th Workshop on Spoken Language Technologies for Under-Resourced Languages (SLTU)*. ISCA, 2018, pp. 66–70.

[15] B. Joshi, B. Bhatta, S. P. Panday, and R. K. Maharjan, "A Novel Deep Learning Based Nepali Speech Recognition," in *Innovations in Electrical and Electronic Engineering*, S. Mekhilef, R. N. Shaw, and P. Siano, Eds. Springer Singapore, 2022, vol. 894, pp. 433–443.

[16] Z. Ghahramani, "An introduction to hidden markov model and bayesian networks," *International Journal of Pattern Recognition and Artificial Intelligence*, 2001.

[17] T. Dutoit, "An introduction to text-to-speech synthesis," *Dordrecht: Kluwer Academic Publishers*, pp. 13, 14, 63,72,179, 196, 1997.

[18] G. Fant, *Acoustic Theory of Speech Production*. Mouton, The Hague, 1960.

[19] K. O'Shea and R. Nash, "An Introduction to Convolutional Neural Networks," 2015.

[20] S. Hochreiter and J. Schmidhuber, "Long short-term memory," *Neural computation*, vol. 9, pp. 1735–80, 12 1997.

[21] K. Cho, B. van Merrienboer, D. Bahdanau, and Y. Bengio, "On the properties of neural machine translation: Encoder-decoder approaches," *CoRR*, vol. abs/1409.1259, 2014.

[22] E. Baral and S. Shrestha, "Large Vocabulary Continuous Speech Recognition for Nepali Language," *International Journal of Signal Processing Systems*, vol. 8, no. 4, pp. 68–73, 2020.

[23] J. Banjara, K. R. Mishra, J. Rathi, K. Karki, and S. Shakya, "Nepali Speech Recognition using CNN and Sequence Models," in *International Conference on Machine Learning and Applied Network Technologies (ICMLANT)*. IEEE, 2020, pp. 1–5.

[24] M. Dhakal, A. Chhetri, A. K. Gupta, P. Lamichhane, S. Pandey, and S. Shakya, "Automatic speech recognition for the Nepali language using CNN, bidirectional LSTM and ResNet," in *International Conference on Inventive Computation Technologies*

[25] K. D. A. Danuwar, K. Badal, S. Karki, S. Titaju, and S. Shrestha, "Nepali Voice-Based Gender Classification Using MFCC and GMM," in *Machine Learning and Computational Intelligence Techniques for Data Engineering*, P. Singh, D. Singh, V. Tiwari, and S. Misra, Eds.  Springer Nature Singapore, 2023, vol. 998, pp. 233–242.

[26] F. He, S.-H. C. Chu, O. Kjartansson, C. Rivera, A. Katanova, A. Gutkin, I. Demirsahin, C. Johny, M. Jansche, S. Sarin, and K. Pipatsrisawat, "Open-source Multi-speaker Speech Corpora for Building Gujarati, Kannada, Malayalam, Marathi, Tamil and Telugu Speech Synthesis Systems," 2020.

[27] K. Li, J. Li, G. Ye, R. Zhao, and Y. Gong, "Towards Code-switching ASR for End-to-end CTC Models," in *International Conference on Acoustics, Speech and Signal Processing (ICASSP)*.  IEEE, 2019.

[28] A. W. Black, "CMU Wilderness Multilingual Speech Dataset," in *International Conference on Acoustics, Speech and Signal Processing (ICASSP)*.  IEEE, 2019, pp. 5971–5975.

[29] R. Ardila, M. Branson, K. Davis, M. Henretty, M. Kohler, J. Meyer, R. Morais, L. Saunders, F. M. Tyers, and G. Weber, "Common Voice: A Massively-Multilingual Speech Corpus," 2020, arXiv:1912.06670.

[30] T. G. Fantaye, J. Yu, and T. T. Hailu, "Investigation of Automatic Speech Recognition Systems via the Multilingual Deep Neural Network Modeling Methods for a Very Low-Resource Language, Chaha," *Journal of Signal and Information Processing*, vol. 11, no. 1, pp. 1–21, 2020.

[31] M. Y. Tachbelie, S. T. Abate, and T. Schultz, "Development of multilingual ASR using globalphone for less-resourced languages: The case of ethiopian languages," in *Proceedings of the Annual Conference of the International Speech Communication Association, INTERSPEECH*, vol. 2020-Octob, 2020, pp. 1032–1036.

[32] B. Thai, R. Jimerson, D. Arcoraci, E. Prud'hommeaux, and R. Ptucha, "Synthetic Data Augmentation for Improving Low-Resource ASR," in *Western New York Image and Signal Processing Workshop (WNYISPW)*.  IEEE, 2019, pp. 1–9.

[33] Z. Tuske, P. Golik, D. Nolden, R. Schluter, and H. Ney, "Data Augmentation, Feature Combination, and Multilingual Neural Networks to Improve ASR and KWS Performance for Low-resource Languages," in *Proceedings of the Annual Conference of the International Speech Communication Association, INTERSPEECH*, 2014.

[34] D.-C. Lyu, T.-P. Tan, E.-S. Chng, and H. Li, "An Analysis of a Mandarin-English Code-switching Speech Corpus: SEAME," 2010.

[35] H. Inaguma, J. Cho, M. K. Baskar, T. Kawahara, and S. Watanabe, "Transfer learning of language-independent end-to-end ASR with language model fusion," in *International Conference on Acoustics, Speech and Signal Processing (ICASSP)*.  IEEE, 2019.

[36] J. Kunze, L. Kirsch, I. Kurenkov, A. Krug, J. Johannsmeier, and S. Stober, "Transfer Learning for Speech Recognition on a Budget," in *Proc. of Workshop on Representation Learning for NLP*, 2017.